\documentclass[aps,preprint]{revtex4}
\usepackage{graphicx,amsmath,amssymb}

\newcommand{\bea}{\begin{eqnarray}}
\newcommand{\eea}{\end{eqnarray}}
\newcommand{\be}{\begin{equation}}
\newcommand{\ee}{\end{equation}}

\begin{document}
\begin{titlepage}
\title{Signatures of dynamic heterogeneity from a
generalized hydrodynamic model of the supercooled liquid}

\author{Bhaskar Sen Gupta and Shankar P. Das}
\affiliation{School of Physical Sciences,\\
Jawaharlal Nehru University,\\
New Delhi 110067, India.}

\setcounter{equation}{0}

\begin{abstract}
In this paper we study the four point correlation function $\chi_4$
of collective density fluctuations in a nonequilibrium liquid. The
equilibration is controlled by a modified stretched exponential
behavior ( $\exp [-{(t_\mathrm{w}/\tau)}^\beta]$ ) having the
relaxation time $\tau$ dependent on the aging time $t_\mathrm{w}$.
Similar aging behavior has been seen experimentally in supercooled
liquids. The basic equations of fluctuating nonlinear hydrodynamics
(FNH) are solved here numerically to obtain $\chi_4$ for equilibrium
and non equilibrium states. We also identify a dynamic length scale
$\xi$ from the equilibrated function. $\xi(T)$ grows with fall of
temperature $T$. From a broader perspective, we demonstrate here
that the characteristic signatures of dynamical heterogeneities in a
supercooled liquid, observed previously in computer simulations of
the dynamics of a small number of particles, are also present in the
coarse grained equations of generalized hydrodynamics.
\end{abstract}


\maketitle
\end{titlepage}

A general feature emerging from simulations\cite{onuki,peter} of the
particle dynamics in a liquid is that at a given instant the
atomic motions in different environments in the structurally
disordered system evolve differently. And yet the fluid particles
constantly move and rearrange so that the distinctions between
different spatial environments of the fluid are transient.
Understanding this complex and evolving situation,
generally termed as dynamic heterogeneities\cite{ediger}
is facilitated through the study of the
multi particle correlation functions.  The multi point structure of the
correlation function is useful in probing the cooperative nature of
the dynamics since it involves incorporating the information at two
different spatial points corresponding to two different times
simultaneously. In a number of recent works, a dynamic length scale
\cite{franz-parisi,sharon-par,berth-fpt,szammel}
depicting the strongly correlated nature of the supercooled
liquid dynamics has been obtained analyzing a four point correlation
function\cite{chandan-epl}.
The different types of four point functions which
have been studied in this respect involve some distinct property of
the fluid\cite{jcp2,hca,reich-imct,flenner,loidl1} like mobility or
density of a tagged particle. In the present paper we
study the dynamics in terms of that of the set
$\{\rho({\bf x},t),{\bf g}({\bf x},t)\}$ respectively
denoting the local densities of mass and momentum of the fluid.
The nature of decay in the fluctuations of these conserved fields
is also the focus of the microscopic theory, termed as
as the mode coupling theory (MCT) for the slow dynamics
in a supercooled liquid.

We compute the time dependent correlation function $\chi_4(q_m,t)$
(to be defined below) involving the collective densities $\rho({\bf
x},t)$ at four points. Here $q_m$ corresponds to
the first maximum  of structure factor. The four point function
develops a sharp peak at a time $t=t_\mathrm{p}$ (say) and eventually decays
out at larger times.
The dynamic length $\xi(T)$, identified from analyzing
\cite{cdg-PRL} the four point function
$\chi_4(q,t_\mathrm{p})$, grows roughly by a factor of three over the
corresponding temperature range. The quantity $\chi_0 \equiv
\chi_4(0,t_\mathrm{p})$ grows as $\xi^{(2-\eta)}$ with the correlation length
$\xi$ with the exponent $2-\eta=2.1$. In
the non equilibrium state without time translation invariance
\cite{kob_barrat}, the four point function $\chi_4(t,t_\mathrm{w})$
for several different values of the waiting (aging)
time $t_\mathrm{w}$, is seen to overlap in the $\alpha$-relaxation regime.
The corresponding frequency transforms $\chi_4(\omega,t_\mathrm{w})$
collapse on a modified Kohlrausch-Williams-Watts (MKWW) relaxation
curve  relaxation time $\tau(t_\mathrm{w})$ dependent on $t_\mathrm{w}$. This
is similar to the behavior seen in two point correlations
\cite{Loidl,bhaskar}.





We consider the product $F(q;t,t_\mathrm{w}) \equiv
\delta\rho(q,t+t_\mathrm{w})\delta\rho(q,t_\mathrm{w})$ of the
fourier transform of the density fluctuations $\delta\rho({\bf
x},t)=\rho({\bf x},t)-\rho_0$ corresponding to wave vector $q$ at
times $t+t_\mathrm{w}$ and $t_\mathrm{w}$ respectively. In the
following $t_\mathrm{w}$ will be referred to as the waiting or aging
time. The normalized two point function is the noise averaged
quantity ${C}(q,t+t_\mathrm{w},t_\mathrm{w})
={\langle}F(q;t,t_\mathrm{w}){\rangle}/
{\langle}F(q,t_\mathrm{w},t_\mathrm{w}){\rangle}$. The long time
limit of the equilibrium averaged quantity
${<F(q;t,0)>}_{\mathrm{eq}}\equiv{\cal C}(q,t)$ changes
discontinuously at the ergodicity non-ergodicity (ENE) transition of
the MCT. The four point function $\chi_4(q;t,t_\mathrm{w})$
normalized with respect to its initial value is defined as, \be
\label{eq:chi4} \chi_4(q;t,t_\mathrm{w}) = \frac{\langle
F(q;t,t_\mathrm{w})F(-q;t,t_\mathrm{w})\rangle}{{|F(q;0,0)|}^2}~~.
\ee To calculate these time correlation functions we need the
dynamical equations controlling the time evolution of density
fluctuations in the liquid. The stochastic equations of fluctuating
nonlinear hydrodynamics (FNH) for the coarse grained densities
 $\{\rho({\bf x},t),{\bf g}({\bf x},t)\}$ are written in the form :
\bea
\label{cont} &&\frac{\partial\rho}{\partial{t}}
+ {\bf \nabla}.{\bf g} = 0, \\
\label{g_eq} && \frac{\partial g_{i}}{\partial t} +
v_0^2 \{\nabla_i \rho
-\rho \nabla_{i} f({\bf x},t)\} + L^0_{ij} \frac{g_j}{\rho} = \theta_{i},
\eea
where $v_o=1/\sqrt{\beta{m}}$ denotes the thermal speed at temperature $T$.
Here $\theta_i$ denotes the thermal noise which  is
assumed to be gaussian and white. The noise correlation is related to
the bare transport matrix $L^0_{ij}$ through the
standard fluctuation dissipation relation.
The function $f({\bf x},t)$ signify the role of interaction between
the fluid particles and is obtained as a convolution function of the
direct correlation function $c({\bf x})$ and the density fluctuation
$\delta\rho({\bf x},t)$. Hence the fourier transform is obtained as
$f(\textbf{k},t) = n_0c(\textbf{k}) \delta \rho(\textbf{k},t)$
where $n_0$ is the equilibrium number density ($\rho_0=mn_0$).
The slow dynamics of the MCT originates from a feedback mechanism
caused by the density nonlinearities in the eqn. (\ref{g_eq}) of FNH.
We have ignored the convective non linearities in the eqn. (\ref{g_eq})
to focus on the role of the coupling of density fluctuations.

The eqns. (\ref{cont})-(\ref{g_eq}) of FNH  are solved numerically on a cubic lattice
of size $20$ with a grid length $h$. Two inputs are required here.
First, the direct correlation function $c(r)$ related to the
structure of the liquid \cite{input-sk}. Second, the bare transport
coefficients $L_{ij}^0$ defining the noise correlations are chosen such that
the corresponding short time dynamics agrees with computer
simulation data. We consider a system of $N$ particles, each of mass $m$
interacting via Lennard-Jones(LJ) potential of characteristic
length scale $\sigma$. Time is scaled with
$\tau_{0}=(m\sigma^{2}/\epsilon)^{\frac{1}{2}}$
and length with lattice constant $h$.
The thermodynamic state of the fluid is
described in terms of the reduced density
$\rho^*=\rho_{0}\sigma^{3}$ and the reduced $T^{*} =
(k_{B}T)/\epsilon$.
The density fluctuations are saved in selected time
bins. A whole array consisting of the density fluctuations
$\rho({\bf x},t)$ on the cubic lattice ${\bf x}$ are transformed
using fast fourier transform subroutines. From this data the two
point and four point correlation functions are respectively
obtained. Several runs for the dynamic
evolution of the system driven by the noise is considered.
Equilibrium is inferred when time translational invariance of the
correlation function is observed, {\em i.e.}, the two time correlation function
$C(t_\mathrm{w},t+t_\mathrm{w})$ depends on $t$ only. This is attained
at increasingly larger $t_\mathrm{w}$ as the liquid is further supercooled.

We equilibrated the system at average density $\rho^*_0=1.10$ and
temperatures respectively at $T^*=1.0$, $0.8$, $0.7$, $0.6$ and
$0.5$. For even lower temperatures $T^*=0.4$ the system does not
equilibrate within the maximum time limit of computation time. The
ratio of the two characteristic lengths $\sigma/h=4.6$ is kept fixed.
The data for $\rho({\bf x},t)$ and ${\bf g}({\bf x},t)$
at each of the grid points are stored for times at equal
intervals extending up to a maximum time $t_\mathrm{max}$
depending on the temperature $T^*$.
For $T^*=.6$ we have $t_\mathrm{max}$~($2000\tau_0$).
To study the equilibrium correlation functions, we
consider large enough initial times $t_\mathrm{w}$.
Fig. 1 displays the four point function $\chi_4(t)$ obtained
by evaluating RHS of eqn. (\ref{eq:chi4}) for $q=q_m$. From the same
data for the $\rho({\bf x},t)$ the two point equilibrium correlation function
$C(t+t_\mathrm{w},t_\mathrm{w}) \equiv {\cal C}(t)$ is also obtained and shown in the
inset of Fig. 1. The two point function ${\cal C}(t)$ reaches a small plateau
value $f_c=0.87$ at $T^*=.8$. Following the predictions of MCT \cite{rmp} the
power law exponents $a$ and $b$ corresponding  the power law and subsequent
von-Sncheider relaxation are 1.27 and 1.18 respectively. For the four point function
$\chi_4$ the peak height $\chi^P$ is attained at $t=t_\mathrm{p}$ which
grows with supercooling indicating the growth of amorphous cluster size.
We obtain $\chi^P \sim {(T-T_o)}^{-1.2}$ with $T_o=.2$ as shown in Fig. 2.
The growth of $t_\mathrm{p}$ observed with fall of $T$ is not as strong
as that of the $\alpha$-relaxation time
$\tau_\alpha$ over the same temperature change.
The inset of Fig. 2 displays the dependence $\chi^P{\sim}t_\mathrm{p}^\mu$
with the exponent $\mu=0.47$.

The four point correlation function $\chi_4(q,t)$ obtained above is
further analyzed to obtain the dynamic
correlation length $\xi$. In Fig 3 we show the scaling of the peak
height $\chi_4(q,t_\mathrm{p})\equiv\chi^P(q)$ for different values of wave
vector $q$ using the Ornstein-Zernike form which includes the $O(q^4)$
\cite{cdg-PNAS,hca} contribution.
From the wave vector dependent data at a fixed
$T$, the correlation length $\xi(T)$ is obtained . The
inset (a) of Fig. 3 shows in the $\xi(T)$ vs. $T$ plot
that the dynamic correlation length does not  diverge
around the so called MCT transition temperature $T_c$.
By fitting the $\alpha$-relaxation time $\tau_\alpha$
to a power law divergence form we obtain $T_c=.4$
in the present case\cite{bhaskar} of one component
LJ system. The length $\xi(T)$ increases only by a factor of
$3$ which is close to corresponding results
seen in MD simulation of a binary LJ mixture \cite{jcp2}
over a similar temperature range. In the
inset (b) of Fig. 3, plot of the peak height
$\chi_0\equiv\chi_4(q=0,T)$ vs. the
correlation length $\xi$ shows that $\chi_0 \sim
\xi^{(2-\eta)}$ with $2-\eta=2.1$. The corresponding value of
($2-\eta$) from simulation of Ref.\cite{cdg-PRL} is $2.2 - 2.4$.
With a simplified form of the MCT model in terms of density only, summing
a class of ladder diagrams  for the four point functions
\cite{biroli-EPL} however obtains a different prediction $2-\eta=4$.
A key observation from our computation of the
two and the four point correlation functions, using the same
density fluctuation data is that the temperature dependence of the relaxation
time $\tau_\alpha(T)$ (obtained from ${\cal C}(t)$)
differs qualitatively from that of the dynamic length scale $\xi(T)$
(obtained from $\chi_4(t)$). We observe that
the $\tau_\alpha(T)$ tends to diverge around a relatively
higher temperature ($T_c$) while the growth of $\xi(T)$
is appears at best to be linked to an underlying transition at $T_g$ or
$T_K$\cite{adam-gibbs} and not to the MCT transition at $T_c$.

To focus on the nonequilibrium dynamics we study the waiting time
($t_\mathrm{w}$) dependence of the four point function $\chi_4(t,t_\mathrm{w})$
defined in eqn. (\ref{eq:chi4}) for  $t_\mathrm{w}=200,
400, 600, 800, 1000$. The $\chi_4(t)$ in each case
grows to a peak of height $\chi^P(t_\mathrm{w})$ (say) at time
$t=t_\mathrm{p}(t_\mathrm{w})$. This is shown in Fig. 4.
The peak time $t_\mathrm{p}$ grows with $t_\mathrm{w}$ and reaches a maximum at an
intermediate $t_\mathrm{w}$ before equilibrating for even longer waiting
times as shown in the inset of Fig. 4. The peak height $\chi^P$ of
the corresponding $\chi_4(t,t_\mathrm{w})$ increases with $t_\mathrm{w}$, signifying
growing dynamic correlation. A parametric plot of $\chi_4(t,t_\mathrm{w})$ vs.
$C(t,t_\mathrm{w})$ is useful for understanding the evolution of the two
point and four point correlations in the non equilibrium system. The
$\alpha$-relaxation parts of the $\chi_4(t,t_\mathrm{w})$ curves for
different $t_\mathrm{w}$ overlap\cite{sriram-saroj}
with the corresponding two
point function $C(t,t_\mathrm{w})$ being shifted by a $t$ independent part
$f(t_\mathrm{w})$. We plot in Fig. 5 the $\chi_4(t,t_\mathrm{w})$ with respect to the
quantity $\tilde{C}(t,t_\mathrm{w})=C(t,t_\mathrm{w})+f(t_\mathrm{w})$.
The part $f(t_\mathrm{w})$
decays to zero as equilibrium is reached as shown in the inset of
Fig. 5. We transform the $\chi_4(t,t_\mathrm{w})$ with respect to
time $t$  to obtain $\chi_4(\omega,t_\mathrm{w})$ corresponding to frequencies
given by $\omega{\tau_0}$=.0001,.0005,.001, and .01.
The data for all $\omega$ values fit well to the form
\be\label{mkww-pres}
\chi(\omega,t_\mathrm{w})=
\left [ \chi^i(\omega)-\chi^f(\omega) \right ]
g(t_\mathrm{w})+\chi^f(\omega),~~\ee
where $\chi^i(\omega)$ and $\chi^f(\omega)$ respectively denote
the initial and final values of the $\chi_4$ at the corresponding $\omega$.
The relaxation function $g(t_\mathrm{w})$ has limiting values $1$ and $0$
respectively as $t_\mathrm{w}{\rightarrow}0$ and $\infty$.  In the main
Fig. 6 we show how the data for all frequencies at $T^*=.6$
collapse on a single curve (solid line) giving a
{\em frequency independent} $g(t_\mathrm{w})$.
The inset displays the $t_\mathrm{w}$ dependence of the relaxation
time $\tau(t_\mathrm{w})$ characterizing the MKWW form
of $g(t_\mathrm{w})$. The relaxation time
$\tau(t_\mathrm{w})$ increases with $t_\mathrm{w}$ implying that aging
slows down at longer waiting time $t_\mathrm{w}$. This is similar to the
observed behavior\cite{Loidl,bhaskar}
with respect to the two point functions (dashed line in the main figure)
obtained from experimental data.
However for the four point functions the time $t_\mathrm{w}$ to reach
saturation in $\tau(t_\mathrm{w})$
is much longer than that for two point case (see inset of Fig. 6).

We have demonstrated here that the appearance of a growing peak in
the four point correlation function $\chi_4(t)$ is a general feature
of the dynamics of the supercooled liquid and it
follows from the basic equations of generalized hydrodynamics
 signifying conservation laws.
This holds even if the two step process ( power law and von-Schneider law)
predicted in the simple MCT\cite{rmp} is not very clearly visible in
the relaxation of two point correlation function ${\cal C}(t)$.
Indeed for the simple LJ system considered here ${\cal C}(t)$
hardly shows any plateau so as to justify a two step relaxation
process. The same density fluctuation data obtains the
prominent peak in the four point function $\chi_4(t)$ growing
with fall of temperature. At a quantitative level, however the
the results for $\chi_4$ obtained from the present work differ from the
predictions of a simplified MCT model which involves an ideal ENE
transition. This is perhaps not unexpected given the fact that the
oversimplified treatment of MCT gets modified in the extended
MCT\cite{das-mazenko} when the implications of the $1/\rho$
nonlinearities are taken into account. From a wider perspective
what is perhaps more relevant\cite{berth-garnier} is that the
general feature of dynamical heterogeneities follow from the basic
equations of FNH which are also the starting point of
the MCT. BSG acknowledges CSIR, India for financial support.
SPD acknowledges support under grant 2011/37P/47/BRNS.


\begin{figure}[!htb]\centering
\includegraphics[width=7cm]{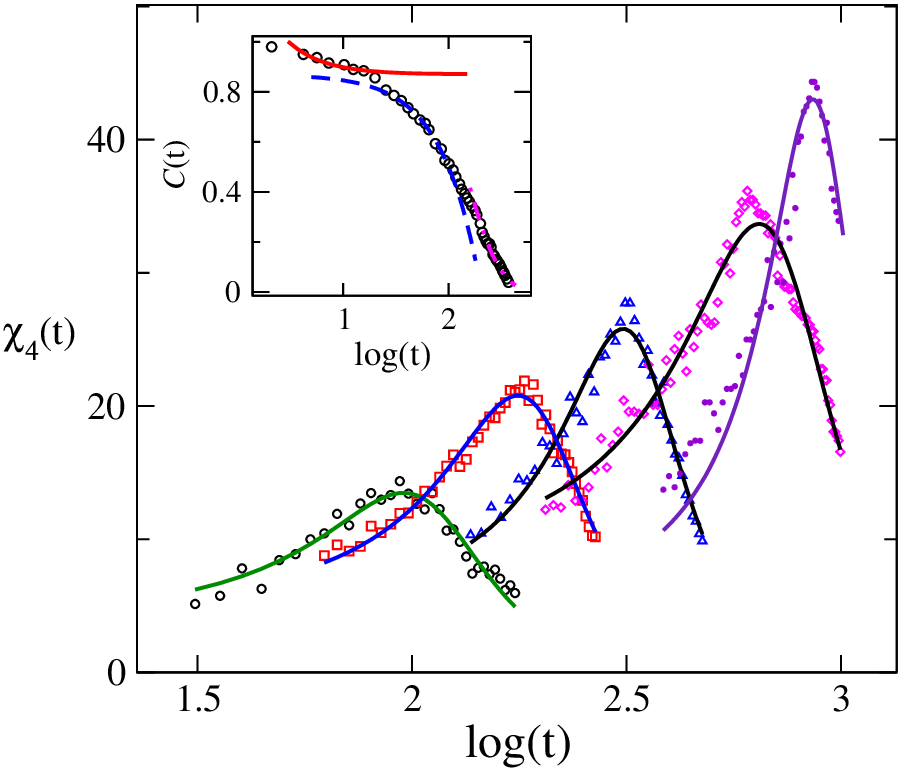} 
\caption{The normalized four point functions $\chi_4(t)$ at $q=q_m$
vs. time $t$ at ${\rho^*_0}=1.10$ and $T^*=1.0$(circles), $0.8$(squares),
$0.7$(triangles), $0.6$(diamonds), $0.5$(stars). Solid lines are the best
fit curves of Lorentzian form. Inset shows the two point function at
${\rho^*_0}=1.10$ and $T^*=0.8$. The solid and dashed
curves indicate the respective power law fits predicted in MCT.}
\label{fig1}
\end{figure}

\begin{figure}[!htb]\centering
\includegraphics[width=7cm]{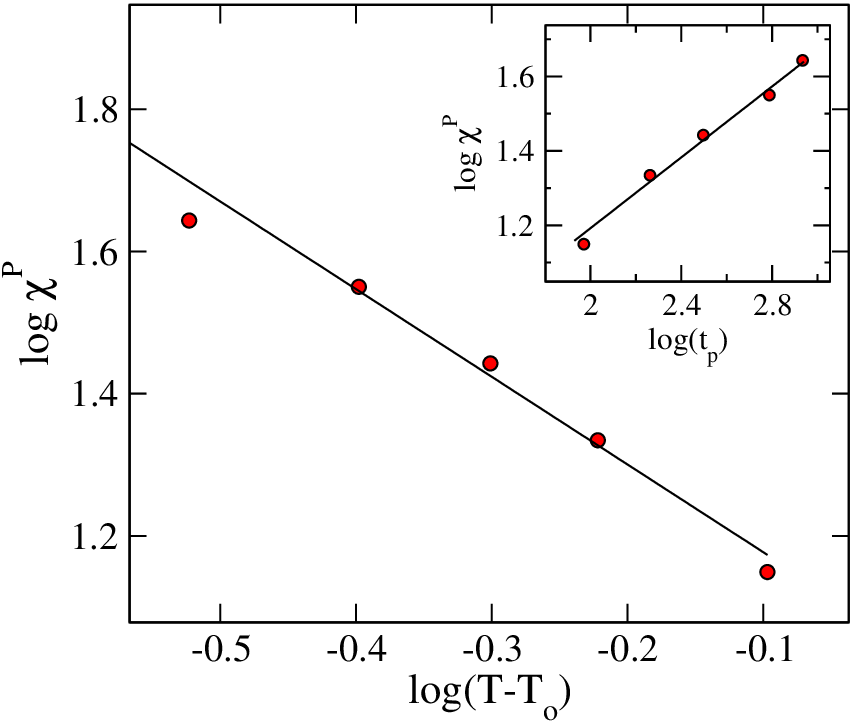} 
\caption{The peak $\chi^P$ of $\chi_4(t)$ at $t=t_\mathrm{p}$
appear to diverge around $T_o=0.2$ with exponent $\alpha=1.21$. Inset shows
$\chi^P{\sim}t_\mathrm{p}^\mu$ behavior with the exponent $\mu=0.47$.} \label{fig2}
\end{figure}

\begin{figure}[!htb]\centering
\includegraphics[width=7cm]{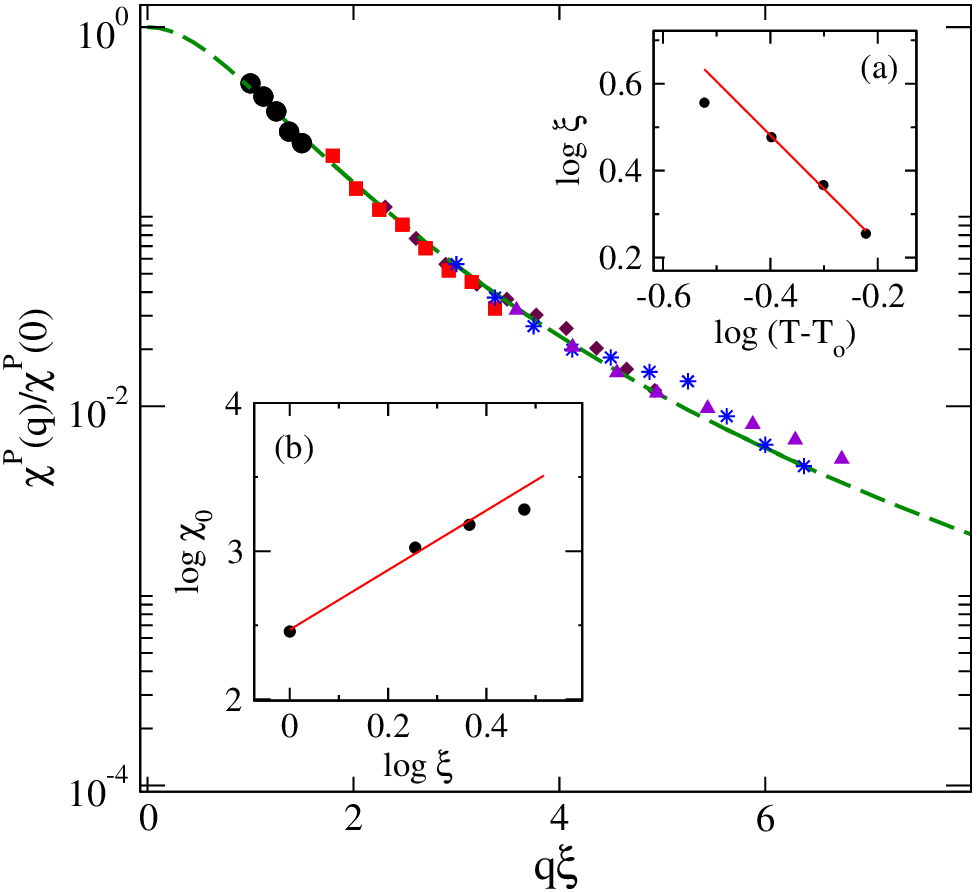} 
\caption{The normalized $\chi^P(q)/\chi^P(0)$ for density
$\rho^*=1.10$ and different temperatures $T^*$=1.0(circles),
0.8(squares),0.7(diamonds),0.6(triangels), and 0.5 (stars) plotted
with corresponding $q\xi(T)$. Dashed line is the best fit to an
Ornstein-Zernike form (see text). Inset (a) shows
divergence $\xi$ around the $T_o=0.2$ and exponent $1.4$.
Inset (b) shows $\chi^P(q=0)\equiv{\chi_0} \sim
\xi^{(2-\eta)}$ with $2-\eta=2.1$.} \label{fig3}
\end{figure}

\begin{figure}[!htb]\centering
\includegraphics[width=7cm]{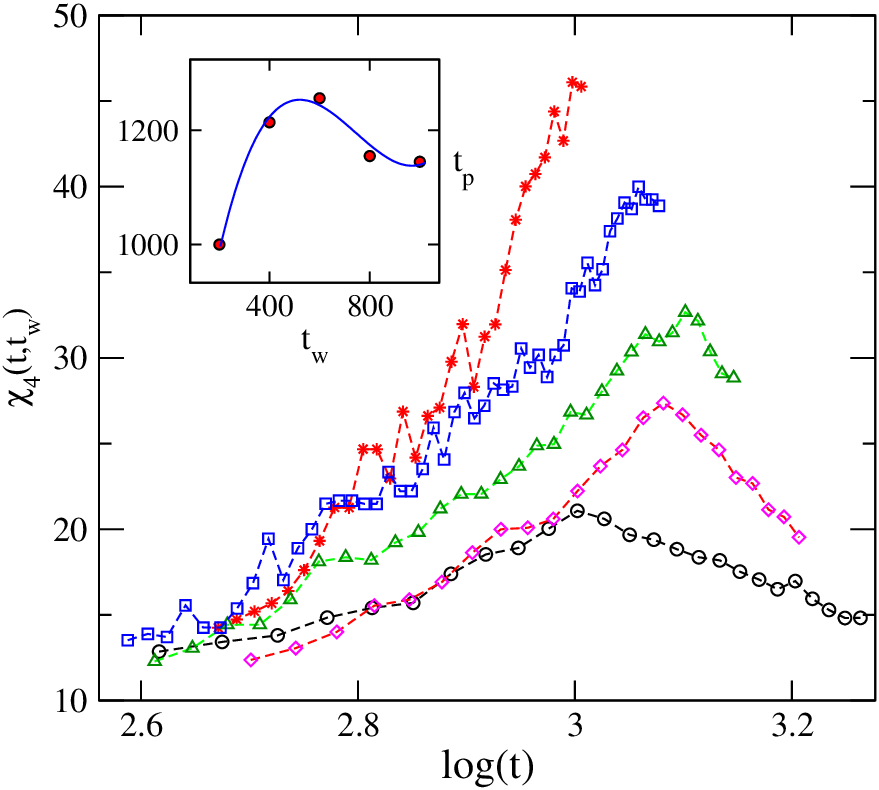} 
\caption{The non equilibrium $\chi_4(t,t_\mathrm{w})$ vs. $t$ for different
values of the waiting time $t_\mathrm{w}$= 200(circles),
400(diamonds), 600(triangles), 800(squares),
and 1000(stars) corresponding to $T^*$=0.4 and $\rho_0^*=1.1$.
Inset : the peak time $t_\mathrm{p}$ vs. $t_\mathrm{w}$.}
\label{fig4}
\end{figure}

\begin{figure}[!htb]\centering
\includegraphics[width=7cm]{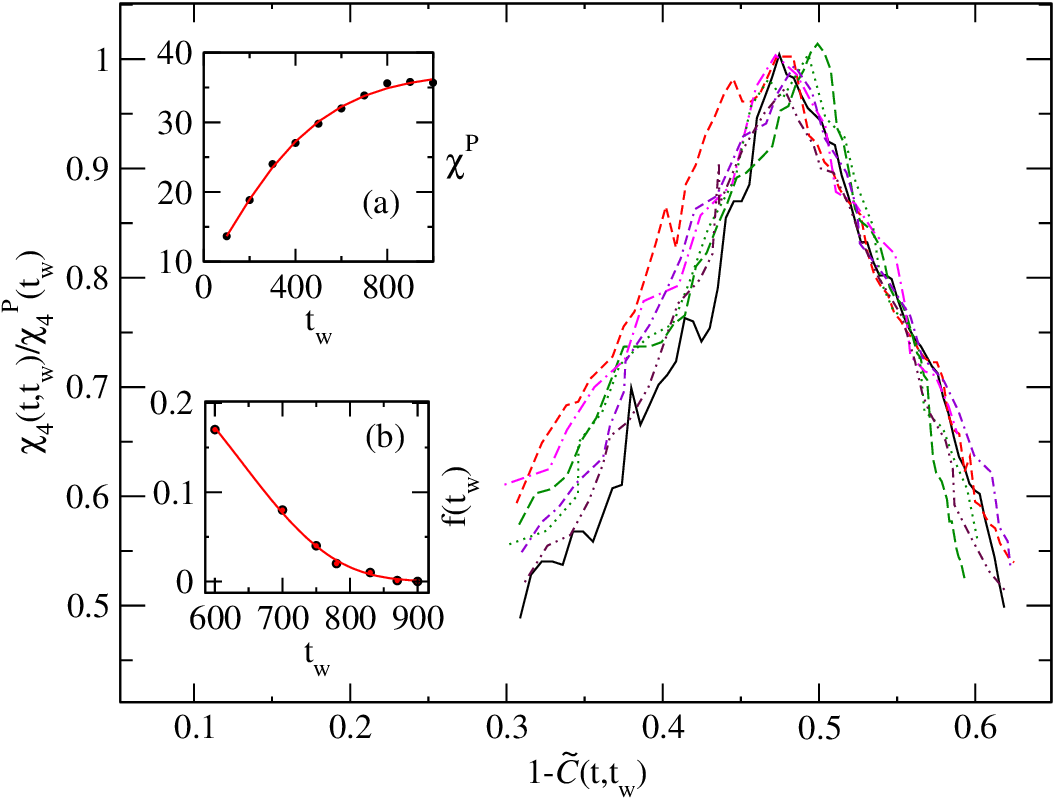} 
\caption{At $T^*=0.6$ and $\rho_0^*=1.1$, parametric plot of
normalized $\chi_4(t,t_\mathrm{w})/\chi^P(t_\mathrm{w})$ vs.
$1-\tilde{C}(t,t_\mathrm{w})$ (see text). Equilibration with waiting
time $t_\mathrm{w}$ : Inset (a) the peak value $\chi^P$; (b)
$f(t_\mathrm{w})$ (defined in text).} \label{fig5}
\end{figure}

\begin{figure}[!htb] \centering
\includegraphics[width=7cm]{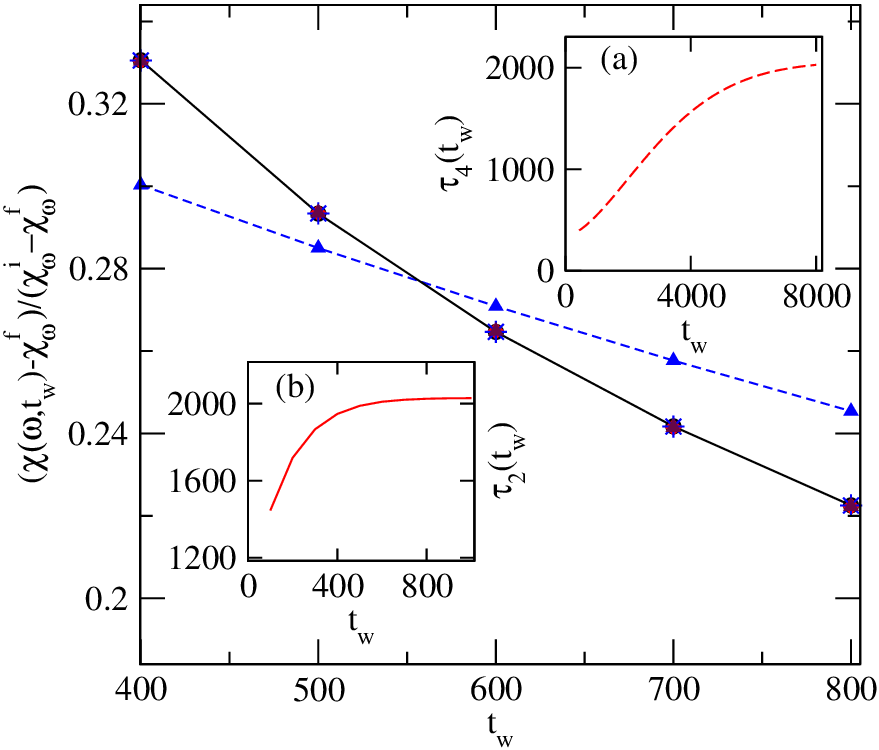} 
\caption{Data collapse (solid line) on the scaling function
$g(t_\mathrm{w})$ (defined in the text) vs. $t_\mathrm{w}$
corresponding to four different frequencies $\omega\tau_0$==
0.0001(circle), 0.0005(diamond), 0.001(triangle), and 0.01(star) at
$T^*=0.6$ and $\rho_0^*=1.1$. Scaling function corresponding to two
point functions (dashed line). The $t_\mathrm{w}$ dependence of
relaxation times of MKWW scaling functions: Inset (a)
$\tau_4(t_\mathrm{w})$ for four point functions; Inset (b)
$\tau_2(t_\mathrm{w})$ for two point functions.} \label{fig6}
\end{figure}

\end{document}